\documentclass[11pt]{article}

\usepackage[T1]{fontenc}
\usepackage[utf8]{inputenc}
\usepackage{lmodern}
\usepackage{microtype}
\usepackage[margin=0.6in]{geometry}
\usepackage{graphicx}
\usepackage{caption}
\usepackage{amsmath,amssymb}
\usepackage{anyfontsize}
\usepackage{colortbl}
\usepackage{xcolor}
\usepackage{steinmetz}
\usepackage{cite}
\usepackage[hidelinks]{hyperref}

\graphicspath{{./images/}}

\title{Fully Fiber-Integrated 3D-Printed Probes for Plug-and-Play Endoscopic Optical Coherence Tomography}
\author{%
Yihan Wang\textsuperscript{1},
Ruilin You\textsuperscript{1},
Jiabin Chen\textsuperscript{1},
Bofan Song\textsuperscript{1},
Zien Feng\textsuperscript{1},\\
Paula Patricia Villarreal\textsuperscript{2},
Gracie Vargas\textsuperscript{2}, and
Rongguang Liang\textsuperscript{1}\thanks{Corresponding author: \href{mailto:rliang@optics.arizona.edu}{rliang@optics.arizona.edu}}\\[0.5em]
\small \textsuperscript{1}Wyant College of Optical Sciences, University of Arizona, Tucson, Arizona 85721, USA\\
\small \textsuperscript{2}The Institute for Translational Sciences, University of Texas Medical Branch,\\
\small Galveston, Texas 77555, USA
}
\date{}

\begin{document}

\maketitle

\begin{abstract}
Miniature probes extend optical coherence tomography (OCT) into lumens and other confined spaces, but conventional implementations often rely on multiple distal components, fiber processing, and probe-specific interferometer matching. Here, a fully fiber-integrated OCT (F2I-OCT) architecture is demonstrated in which two-photon microfabrication defines not only the terminal imaging optic, but the complete distal optical and interferometric architecture within a single fiber-mounted element. Beam expansion, side-view redirection, common-path reference generation, and terminal imaging are physically integrated while remaining independently designable. This architecture transfers complexity from component fabrication and interferometer matching into three-dimensional optical design, reducing assembly to a print-and-bond process and enabling plug-and-play exchange on the same OCT platform. Terminal optics can be adapted to different working distances, surrounding media, and wavefront transformations without redesigning the upstream architecture. Twenty assembled probes exhibit returned-reference and side-viewing-output power standard deviations of $0.149~\mathrm{dB}$ and $0.071~\mathrm{dB}$, respectively. The system achieves $93.1~\mathrm{dB}$ sensitivity and enables ex vivo imaging of airway and dental structures. The combination of a common plug-and-play architecture with broad optical design freedom provides a versatile platform for endoscopic and confined-space imaging across diverse biomedical and technical environments.
\end{abstract}


\section{Introduction}\label{sec:introduction}

Optical coherence tomography (OCT) provides depth-resolved, label-free imaging of tissue microstructure and has become an important modality for optical biopsy\cite{bouma2022optical,gora2017endoscopic}. Extending OCT into lumens, cavities, and other confined spaces relies on fiber-scale probes that place the imaging optics at the distal end of the system. Such probes enable applications ranging from intravascular imaging in liquid-filled lumens to small-airway imaging in air-filled lumens, as well as image-guided intervention and inspection of confined engineered structures\cite{gora2017endoscopic,yuan2022direct,wang2024automatic,he2023robotic}. Their broader deployment, however, places increasingly stringent demands on the distal optics, which must combine miniaturization with reproducible fabrication, straightforward exchange, and adaptability to different imaging geometries and operating environments\cite{li2020ultrathin,xu2024liquid,chantawannakul2026toward}.

Conventional endoscopic OCT probes typically combine multiple distal optical components with a separate reference arm that requires delay matching, dispersion balancing, and polarization control\cite{gora2017endoscopic,wang2007superachromatic,yuan2017superachromatic}. Their fabrication also commonly relies on fiber processing, precision cleaving, and sequential component alignment\cite{wang2007superachromatic,yuan2017superachromatic,li2020ultrathin,xu2024liquid}. Common-path OCT partially reduces the system-level burden by eliminating the need for a separately matched reference arm, but it does not remove the fabrication and assembly requirements of the distal probe\cite{sharma2005allfiber,sharma2007common,tumlinson2006endoscope}. In interface-based implementations, the reference level and zero-delay position are further constrained by the properties and physical location of the reference interface\cite{park2012double,wang2022truncated}. These limitations point to the need for a distal architecture that reduces both fabrication and operating complexity while retaining independent control over its principal optical functions.

Miniaturized OCT probes have increasingly incorporated advanced distal optics to control focusing, aberration, and wavefront structure within fiber-scale form factors, including nano-optic and metasurface implementations\cite{pahlevaninezhad2018nano,pahlevaninezhad2022metasurface}. Two-photon polymerization (2PP) has further advanced this design space by enabling freeform micro-optics to be fabricated directly at the distal end of optical fibers\cite{li2018twophoton,li2020ultrathin}. In endoscopic OCT, this capability has enabled compact beam redirection, aberration correction, focusing, and wavefront shaping that are difficult to realize with conventional micro-optical assembly\cite{li2018twophoton,li2020ultrathin,ruchka2025printed,gu2024computer}. To date, however, the design freedom of 2PP has been exploited primarily at the level of individual optical surfaces or terminal imaging functions\cite{li2018twophoton,li2020ultrathin,li2022lensinlens,ruchka2025printed,chen2025low}. Beam expansion, interferometric reference generation, and other upstream functions generally remain distributed among separately fabricated fiber sections or external system components\cite{li2020ultrathin,ruchka2025printed,xu2024liquid}. An architecture-level approach instead extends additive microfabrication from individual distal micro-optics to the complete distal optical and interferometric architecture, allowing multiple functions to be co-designed within a common three-dimensional structure.

Here, a fully fiber-integrated OCT (F2I-OCT) architecture is demonstrated in which a single 2PP-printed element defines the complete distal interferometric architecture. Beam expansion, side-view redirection, common-path reference generation, and terminal imaging are integrated within one fiber-mounted optical body while remaining functionally decoupled at the design level. This integration reduces probe fabrication to a print-and-bond process and allows the probe-defined reference to support plug-and-play exchange on the same OCT system without probe-specific interferometer adjustment. The terminal optics can be adapted to different imaging geometries, operating environments, and wavefront-shaping requirements without redesigning the upstream expansion and reference architecture. Reproducible assembly is demonstrated across twenty consecutive probes, together with sensitivity above 93~dB and ex vivo imaging in biological structures. More broadly, F2I-OCT shifts the role of two-photon printing from fabricating individual distal micro-optics to defining complete fiber-scale optical architectures, transferring complexity from component assembly and interferometer adjustment into monolithic three-dimensional design.
 
 \section{Architecture and Design}\label{sec:architecture}

 \subsection{Fully fiber-integrated architecture}\label{subsec:architecture}

\begin{figure}[htbp]
\centering
\includegraphics[width=0.9\textwidth]{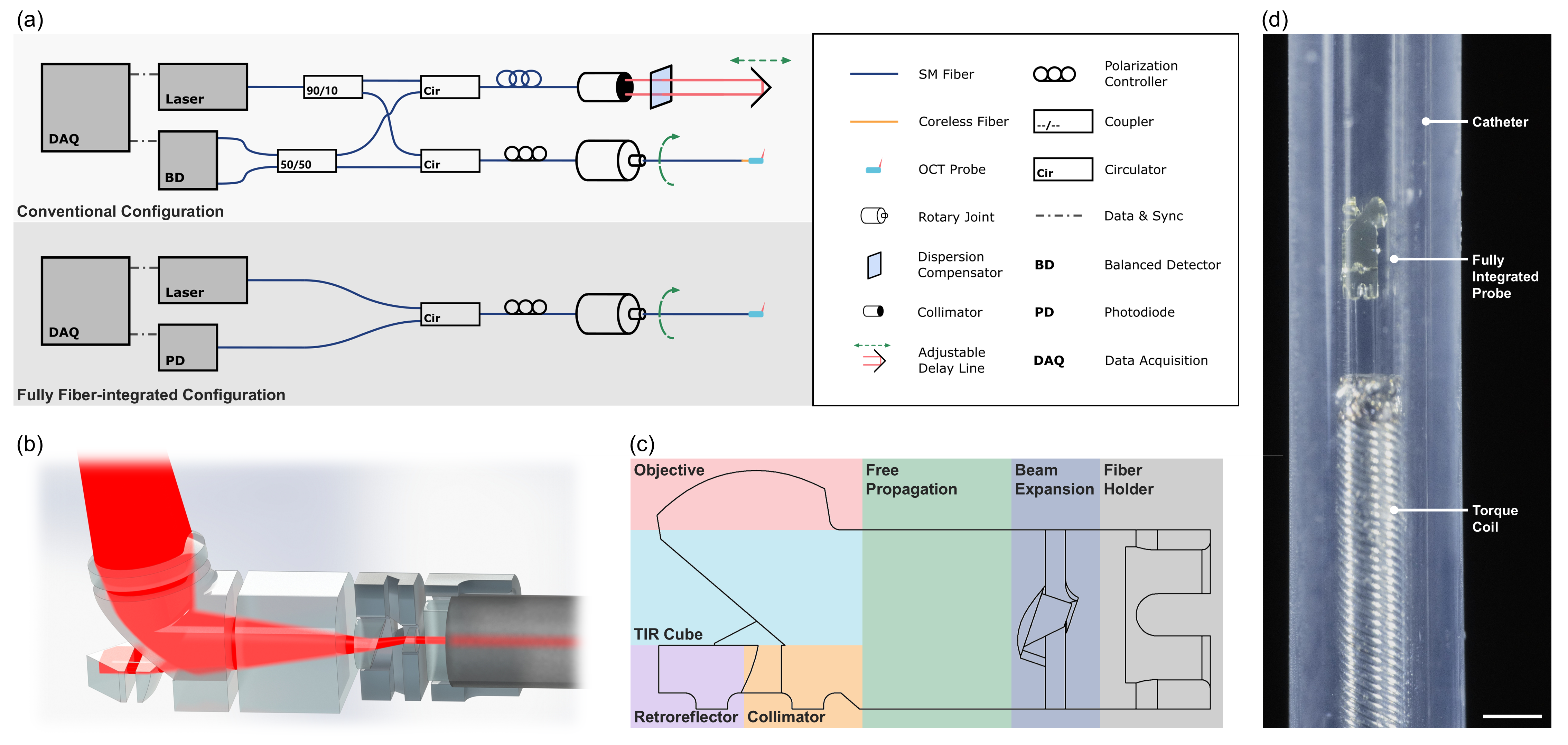}
\caption{Fully fiber-integrated common-path OCT architecture.
(a) Conventional dual-arm endoscopic OCT (top) and the F2I-OCT architecture (bottom), in which the distal optical and reference functions are integrated within a single 2PP-printed element.
(b) Three-dimensional rendering of the probe and optical path. The expanded wavefront is divided between the side-view imaging path and the embedded common-path reference.
(c) Cross-sectional schematic showing the functional optical surfaces.
(d) Assembled probe inside the catheter sheath. Scale bar, $300~\mu$m.
}
\label{fig:concept}
\end{figure}

Figure~\ref{fig:concept}a contrasts the component-based configuration commonly used in endoscopic OCT with the F2I-OCT architecture developed here. Beam expansion, side-view redirection, terminal imaging, and common-path reference generation are integrated within a single 2PP-printed element mounted on a cleaved single-mode delivery fiber. Their relative positions and optical functions are defined by a common three-dimensional geometry rather than by sequential assembly of discrete components. Because the sample and reference fields share the same delivery path, the reference level and zero-delay position are defined by the probe itself. The distal probe is therefore reduced to a printed optical architecture that requires only fiber--element bonding for final assembly.

The optical path is shown in Fig.~\ref{fig:concept}b,c. Light emerging from the delivery fiber first passes through an in-element beam-expansion stage before reaching a tilted total-internal-reflection surface. Most of the expanded field is redirected toward the side-view imaging path, whereas a designed sub-aperture transmits a selected portion to an embedded collimator and retroreflector. The returned field forms the common-path reference and propagates back through the same delivery fiber as the sample signal.

Importantly, physical integration does not couple the principal design degrees of freedom. The expansion stage defines the incident field, the wavefront-dividing sub-aperture controls the reference level, the folded reference geometry sets the reference delay, and the terminal optics determine the imaging function. The terminal design can therefore be adapted to different working distances, immersion conditions, and wavefront-shaping requirements without modifying the upstream expansion and reference architecture.

\subsection{In-element beam expansion}\label{subsec:expansion}

\begin{figure}[htbp]
\centering
\includegraphics[width=0.9\textwidth]{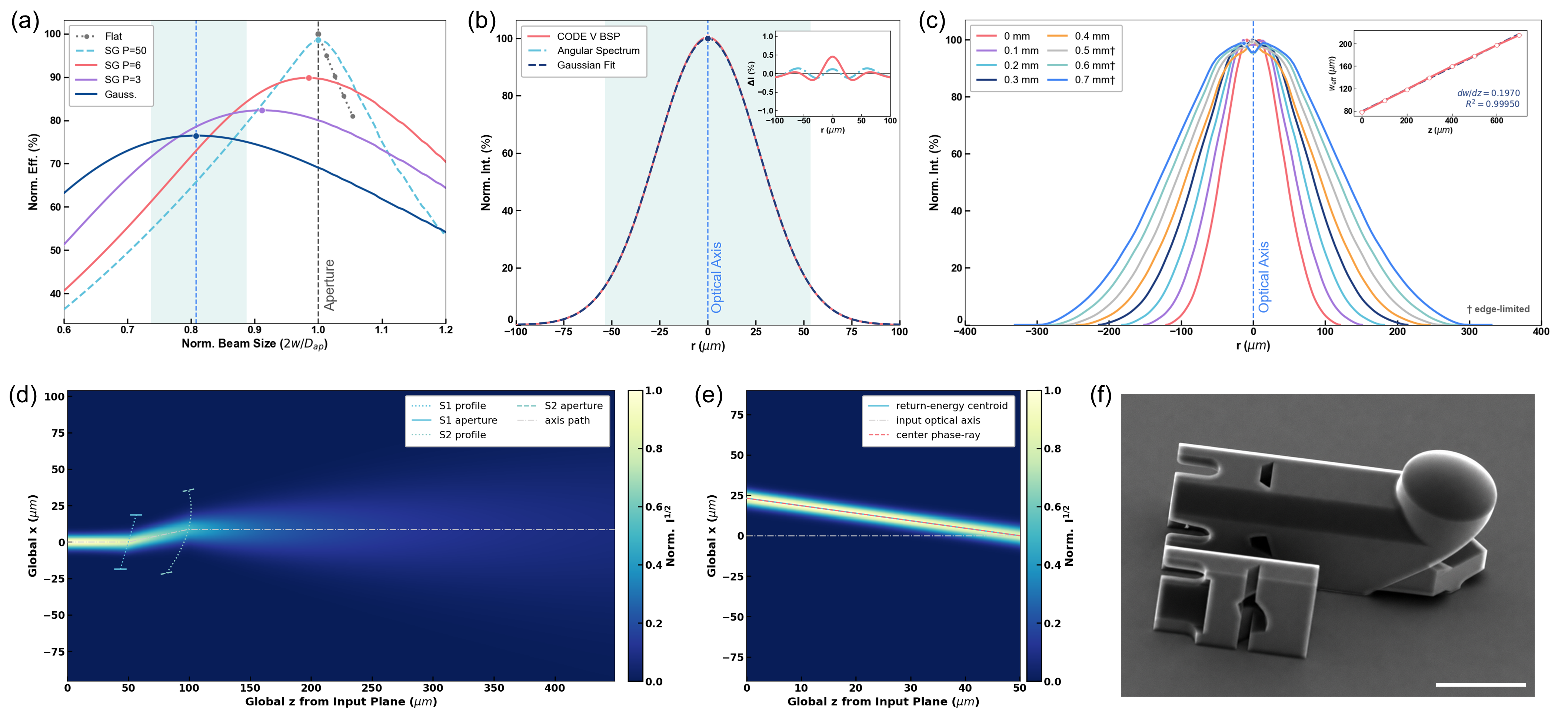}
\caption{Design and characterization of the in-element beam expansion.
(a) Calculated round-trip efficiency versus aperture-normalized filling factor $\chi=2w/D_{\mathrm{ap}}$ for Gaussian, super-Gaussian, and flat-top beam profiles. The Gaussian optimum occurs at $\chi\simeq0.81$, with efficiency above $75\%$ over $\chi\simeq0.74$--$0.89$.
(b) Simulated output intensity profile at the exit of the expansion stage, comparing angular-spectrum propagation, CODE~V beam synthesis, and a Gaussian fit. Inset, residuals relative to the Gaussian fit.
(c) Measured radial beam profiles over a $0.7~\mathrm{mm}$ propagation range. Inset, effective beam radius versus propagation distance, yielding $\mathrm{d}w/\mathrm{d}z=0.1970$ ($R^{2}=0.9995$).
(d) Simulated field propagation through the two refractive surfaces and internal air gap.
(e) Back-propagation of the first-surface Fresnel reflection, showing its displacement from the fiber optical axis and suppression by spatial-mode mismatch.
(f) Scanning electron micrograph of a complete printed element (upper) and a separately printed cross-section of the expansion region (lower). Scale bar, $150~\mu$m.}
\label{fig:beamExpansion}
\end{figure}

Miniature endoscopic OCT probes commonly expand the fiber mode by fusion-splicing a short coreless-fiber segment to the delivery fiber\cite{yuan2017superachromatic,li2020ultrathin,xu2024liquid,ruchka2025printed}. The resulting beam geometry depends on the splice position, cleave position, and segment length, introducing fabrication-dependent variations into each probe. In F2I-OCT, beam expansion is instead defined by two refractive surfaces within the printed element. This converts the fiber-to-beam transformation into a geometry-defined optical function while registering the expanded field to the downstream reference and imaging surfaces within the same three-dimensional structure.

The target beam size was selected from the trade-off among transverse resolution, collection efficiency, and sidelobe level under the fixed probe aperture. The aperture-normalized filling factor is defined as $\chi=2w/D_{\mathrm{ap}}$, where $w$ is the $1/e^{2}$ intensity radius and $D_{\mathrm{ap}}$ is the clear-aperture diameter. For a Gaussian beam, the round-trip efficiency accounting for forward aperture transmission and reciprocal mode overlap is
\begin{equation}
    \eta_{\mathrm{G}}(\chi)=
    2\chi^{2}
    \left(1-e^{-1/\chi^{2}}\right)^{2}
    \left(1-e^{-2/\chi^{2}}\right),
    \label{eq:gaussian_aperture_efficiency}
\end{equation}
which reaches approximately $76.5\%$ at $\chi\simeq0.81$ and remains above $75\%$ over $\chi\simeq0.74$--$0.89$ (Fig.~\ref{fig:beamExpansion}a). Higher-order super-Gaussian and flat-top profiles can provide higher ideal efficiencies and narrower main lobes\cite{han2011flattop}, but their sharper transverse boundaries evolve more strongly during propagation and are more sensitive to aperture truncation and fabrication error. A Gaussian target was therefore selected because it preserves a stable propagation profile and provides a compact analytic field for downstream optical design.

Because the expansion occurs over a distance comparable to the Rayleigh range, the first refractive surface lies in the near field of the fiber mode. The two surfaces were therefore designed directly from the Gaussian amplitude and phase rather than from a far-field approximation. Equal-power annuli of the incident and target fields were mapped between the two interfaces, with the corresponding surface normals determined from vector Snell refraction,
\begin{equation}
\begin{aligned}
    P_j(r)&=1-\exp\!\left(-\frac{2r^2}{w_j^2}\right),
    &P_1(r_1)&=P_2(r_2),\\
    \mathbf{N}&\parallel n_i\mathbf{s}_i-n_t\mathbf{s}_t .
\end{aligned}
    \label{eq:power_mapping_snell}
\end{equation}
where $P_j(r)$ is the encircled power of the Gaussian field, $\mathbf{s}_i$ and $\mathbf{s}_t$ are the incident and transmitted ray directions, and $n_i$ and $n_t$ are the corresponding refractive indices. For $\lambda=1.31~\mu\mathrm{m}$, a polymer refractive index of $1.498$, and an input numerical aperture of $0.091$, the design gives a target divergence of $\mathrm{d}w/\mathrm{d}z=0.1958$. The first surface is positioned $50~\mu\mathrm{m}$ from the fiber input plane and separated from the second surface by a $50~\mu\mathrm{m}$ air gap. At the downstream aperture, the expanded field reaches a $1/e^{2}$ radius of $61~\mu\mathrm{m}$, corresponding to $\chi=0.76$ for the $160~\mu\mathrm{m}$ clear aperture.

The polymer--air interface produces a Fresnel reflection of approximately $3.97\%$, which would otherwise form a strong uncontrolled common-path return. Paired linear tilts are therefore incorporated into the two refractive surfaces to reject this reflection without altering the intended expansion. The first surface deflects the transmitted field within the air gap, while the second restores propagation parallel to the input optical axis (Fig.~\ref{fig:beamExpansion}d). At the same time, the first-surface reflection is directed away from the fiber core. Back-propagation predicts a return-field displacement of approximately $49~\mu\mathrm{m}$ at the fiber input plane, compared with a fiber-mode radius of $4.58~\mu\mathrm{m}$ (Fig.~\ref{fig:beamExpansion}e), strongly suppressing recoupling of the parasitic reflection and leaving the embedded reference branch described in Section~\ref{subsec:commonPath} as the controlled interferometric return.

Numerical propagation confirms that the expansion stage produces the intended Gaussian field. Angular-spectrum propagation and independent CODE~V beam synthesis yield closely matched Gaussian profiles at the output plane (Fig.~\ref{fig:beamExpansion}b). The output wavefront was independently fitted as both a Gaussian beam and a spherical wave, with the inferred source positions agreeing to within $0.5\%$. The expansion stage can therefore be treated as a well-defined equivalent point source for the independently designed reference and terminal imaging modules.

Scanning electron microscopy confirms formation of the designed refractive surfaces, surface tilt, and internal air gap (Fig.~\ref{fig:beamExpansion}f). Experimentally, radial beam profiles measured with an InGaAs camera over a $0.7~\mathrm{mm}$ propagation range remained closely Gaussian, with an RMS deviation below $2.7\%$ of the peak intensity at every measured position (Fig.~\ref{fig:beamExpansion}c). The measured beam radius increased linearly with propagation distance, yielding $\mathrm{d}w/\mathrm{d}z=0.1970$ with $R^{2}=0.9995$, within $0.6\%$ of the designed value of $0.1958$. The in-element expander therefore converts beam expansion from a fiber-processing-dependent fabrication step into a reproducible, geometry-defined optical function while providing a common upstream field for the independently designed reference and terminal imaging functions.

\subsection{Wavefront-division common-path reference}\label{subsec:commonPath}

\begin{figure}[htbp]
\centering
\includegraphics[width=0.9\textwidth]{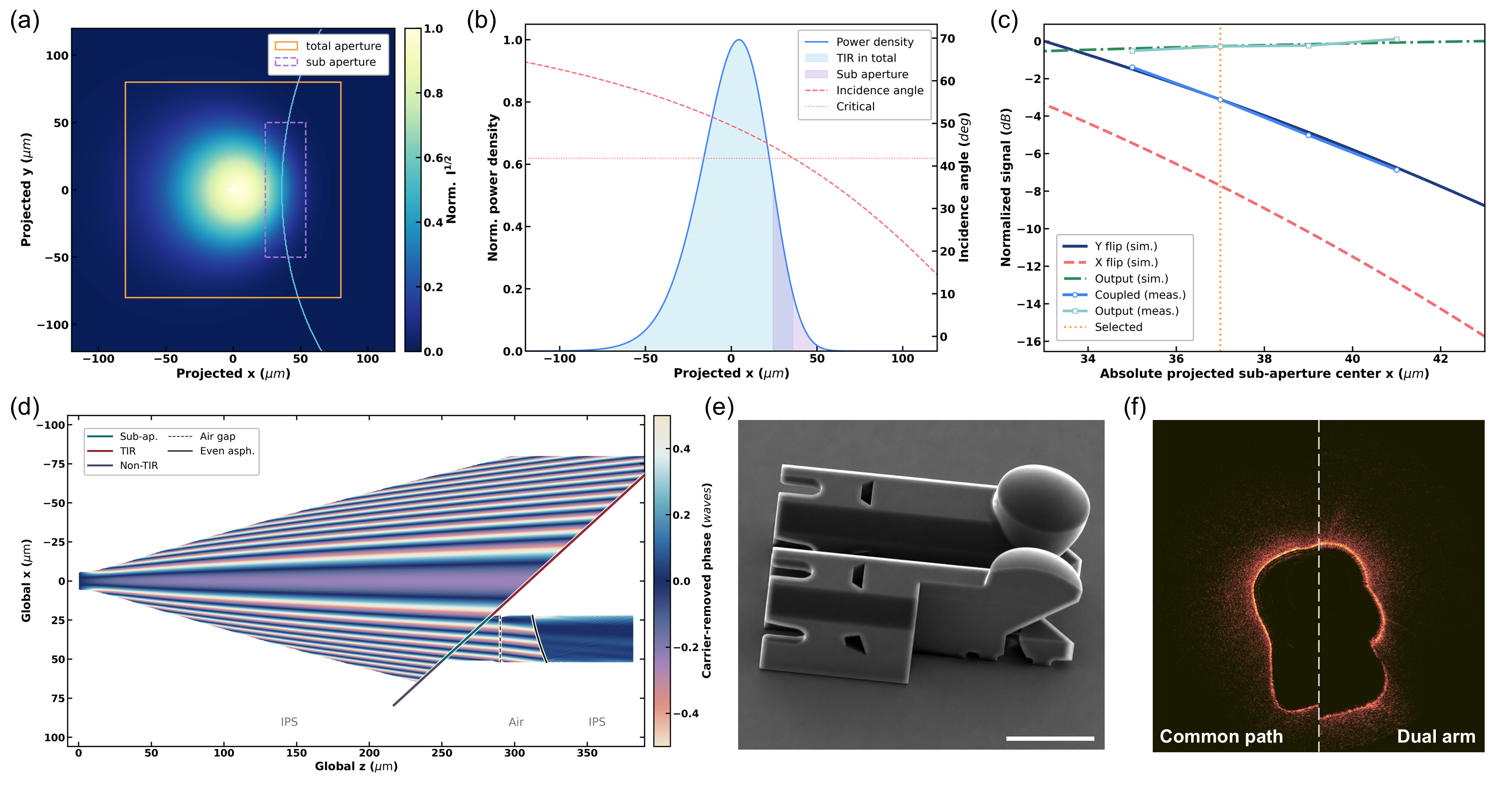}
\caption{Wavefront-division common-path reference.
(a) Expanded Gaussian field projected onto the tilted TIR surface. The clear aperture, reference sub-aperture, critical-angle contour, and optical axis are indicated.
(b) Power density and local incidence angle along the surface tilt direction, showing the totally internally reflected imaging region and the sub-critical reference region.
(c) Simulated and measured reference return versus projected sub-aperture position. The transmitted imaging power remains nearly unchanged over the same range, and the dotted line marks the selected design position.
(d) Angular-spectrum propagation through the reference branch, showing collimation of the transmitted sub-aperture field and its return from the embedded retroreflector.
(e) Scanning electron micrographs of a complete printed element and a separately printed cross-section through the TIR and reference regions. Scale bar, $150~\mu$m.
(f) OCT images of an ex vivo dental pulp cavity acquired with the integrated common-path probe (left) and a path-matched dual-arm configuration using the same distal imaging design (right). Neither configuration used dispersion compensation.}
\label{fig:commonPath}
\end{figure}

Common-path interferometry is attractive for endoscopic OCT because the sample and reference fields share the same delivery fiber and much of the distal optical path, suppressing differential path-length, dispersion, and polarization mismatch\cite{vakhtin2003common,sharma2005allfiber,sharma2007common}. The practical benefit is illustrated in Fig.~\ref{fig:commonPath}f. The F2I-OCT probe was compared with a path-matched dual-arm configuration using an otherwise identical distal imaging design but without the printed reference module. The dual-arm reference path was independently matched, and neither configuration used dispersion compensation. Under these conditions, the integrated common-path probe provides sharper structural delineation in an ex vivo dental pulp cavity. Conventional common-path implementations usually derive the reference from an existing optical interface, so the reference intensity follows the interface reflectance and the zero-delay position is tied to its physical location\cite{park2012double,wang2022truncated}. In F2I-OCT, the tilted total-internal-reflection (TIR) surface instead performs wavefront division, routing most of the expanded field toward the side-view imaging path while transmitting a designed sub-aperture to an embedded collimator and retroreflector. The reference level and delay thereby become design parameters of the printed element rather than fixed properties of an existing interface.

The wavefront division follows from the variation of local incidence angle across the tilted TIR surface. For a surface with normal
$\hat{\mathbf{n}}=(\sin\alpha,0,\cos\alpha)$, the local propagation direction of the expanded Gaussian field and the corresponding incidence angle are
\begin{equation}
    \hat{\mathbf{s}}(x,y,z)\propto
    \left(\frac{x}{R(z)},\frac{y}{R(z)},1\right),
    \qquad
    \cos\theta_i=\hat{\mathbf{s}}\cdot\hat{\mathbf{n}},
    \label{eq:local_incidence}
\end{equation}
where $R(z)$ is the Gaussian wavefront radius of curvature and $\theta_{\mathrm{c}}$ is the critical angle. For the $49.5^{\circ}$ surface used here, $\theta_{\mathrm{c}}=41.9^{\circ}$. The incidence angle exceeds the critical angle over most of the expanded aperture, redirecting the main field toward the side-view objective, while it falls below the critical angle near one edge of the beam (Fig.~\ref{fig:commonPath}a,b). A sub-aperture placed within this sub-critical region therefore extracts a controlled portion of the wavefront for reference generation without requiring an additional coating or discrete beam splitter.

The Gaussian field established by the upstream expansion stage makes the reference level predictable at the design stage. The power intercepted by the reference sub-aperture is
\begin{equation}
    P_{\mathrm{sub}}
    =\iint_{A_{\mathrm{sub}}}I_{\mathrm{s}}(x,y)\,\mathrm{d}A ,
    \label{eq:reference_power}
\end{equation}
where $I_{\mathrm{s}}(x,y)$ is the projected Gaussian power density on the TIR surface. The transmitted field is collimated by an even-aspheric surface designed against the equivalent point source defined by the expansion stage, giving an intensity-weighted RMS wavefront error of $0.036\lambda$ in simulation (Fig.~\ref{fig:commonPath}d). The right-angle retroreflector introduces a transverse parity inversion. For the implemented orientation,
\begin{equation}
    E_{\mathrm r}(x,y)
    =\sqrt{\rho_{\mathrm{ref}}}\,
    E(x,2y_{\mathrm c}-y),
    \qquad
    \eta_{\mathrm{ref}}
    \simeq
    \rho_{\mathrm{ref}}
    \left(\frac{P_{\mathrm{sub}}}{P_0}\right)^2 ,
    \label{eq:retro_coupling}
\end{equation}
where $P_0$ is the incident power and $\rho_{\mathrm{ref}}$ is the lumped round-trip power factor of the reference branch excluding the power selection imposed by the sub-aperture. The retroreflector is oriented so that the parity inversion occurs along $y$, where the sub-aperture is centred on the beam axis and the Gaussian field remains symmetric. The returned field therefore retains strong modal overlap with the guided mode. An inversion along $x$ would instead mirror the laterally offset sub-aperture and reduce the coupled reference.

The reference level is tuned directly through the sub-aperture position. Moving the sub-aperture outward intercepts a smaller fraction of the Gaussian field and therefore strongly reduces the returned reference while only weakly affecting the transmitted imaging power. Experimentally, the reference return changed by approximately $5~\mathrm{dB}$ over a $6~\mu\mathrm{m}$ displacement, whereas the imaging output remained within $0.5~\mathrm{dB}$ over the same range (Fig.~\ref{fig:commonPath}c). A projected sub-aperture centre position of $37~\mu\mathrm{m}$ was selected for the probes used in this study. The reference level can therefore be specified through the printed geometry without modifying the upstream expansion stage, terminal imaging optics, or OCT-system settings. More generally, this design variable allows the reference strength to be adapted to the expected sample return, providing stronger references for weakly scattering samples and weaker references for highly reflective targets without altering the underlying interferometric architecture.

The reference delay is independently determined by the folded optical path inside the printed element. From the fiber facet to the retroreflector, the reference branch contains $50~\mu\mathrm{m}$ of air and $390~\mu\mathrm{m}$ of polymer. The resulting optical path length and round-trip delay are
\begin{equation}
    \mathrm{OPL}_{\mathrm{ref}}
    =n_{\mathrm{a}}d_{\mathrm{air}}
    +n_{\mathrm{p}}d_{\mathrm{pol}}
    =634~\mu\mathrm{m},
    \qquad
    \Delta_{\mathrm{rt}}
    =1.27~\mathrm{mm}.
    \label{eq:path_budget}
\end{equation}
This delay places the zero-delay plane close to the rotational axis of the side-view probe, which is advantageous for rotational OCT because the reconstruction origin naturally coincides with the probe axis while strong sheath reflections and their complex-conjugate images are displaced away from the principal tissue region. In interface-derived common-path configurations, changing the reference delay generally requires moving the physical reference interface. Here, the folded reference geometry can instead be adjusted within the printed element, allowing the zero-delay position to be selected independently of the terminal imaging design. Because the folded path can be extended primarily along the fiber axis, additional reference delay can also be introduced with little increase in the radial probe dimension. Moreover, because the optical path difference is encoded directly in the printed CAD geometry rather than established through probe-specific mechanical adjustment, its value is set by the fabrication geometry and material optical path. This geometry-defined reference could provide improved probe-to-probe reproducibility of the zero-delay position and be advantageous for quantitative dimensional measurements in confined-space inspection and other metrological applications\cite{fu2024progress,kahatapitiya2026optical}.

The printed TIR, collimating, and retroreflecting surfaces are shown in Fig.~\ref{fig:commonPath}e. Embedding the reference branch within the printed element allows the fiber--element junction to be fully bonded without retaining an external air gap for reference generation. More importantly, wavefront division occurs upstream of the terminal optics, so redesigning the imaging function does not require redesigning the reference module. Together with the geometry-defined beam expansion in Section~\ref{subsec:expansion}, the reference level, reference delay, and terminal imaging function can therefore be varied independently within the same monolithic architecture. Because the interferometric state is carried by the probe rather than established by an external reference arm, probes with different terminal designs can be exchanged on the same OCT system without probe-specific interferometer matching, providing the architectural basis for the plug-and-play operation demonstrated in Section~\ref{sec:results}.

\section{Results}\label{sec:results}

\begin{figure}[htbp]
\centering
\includegraphics[width=0.9\textwidth]{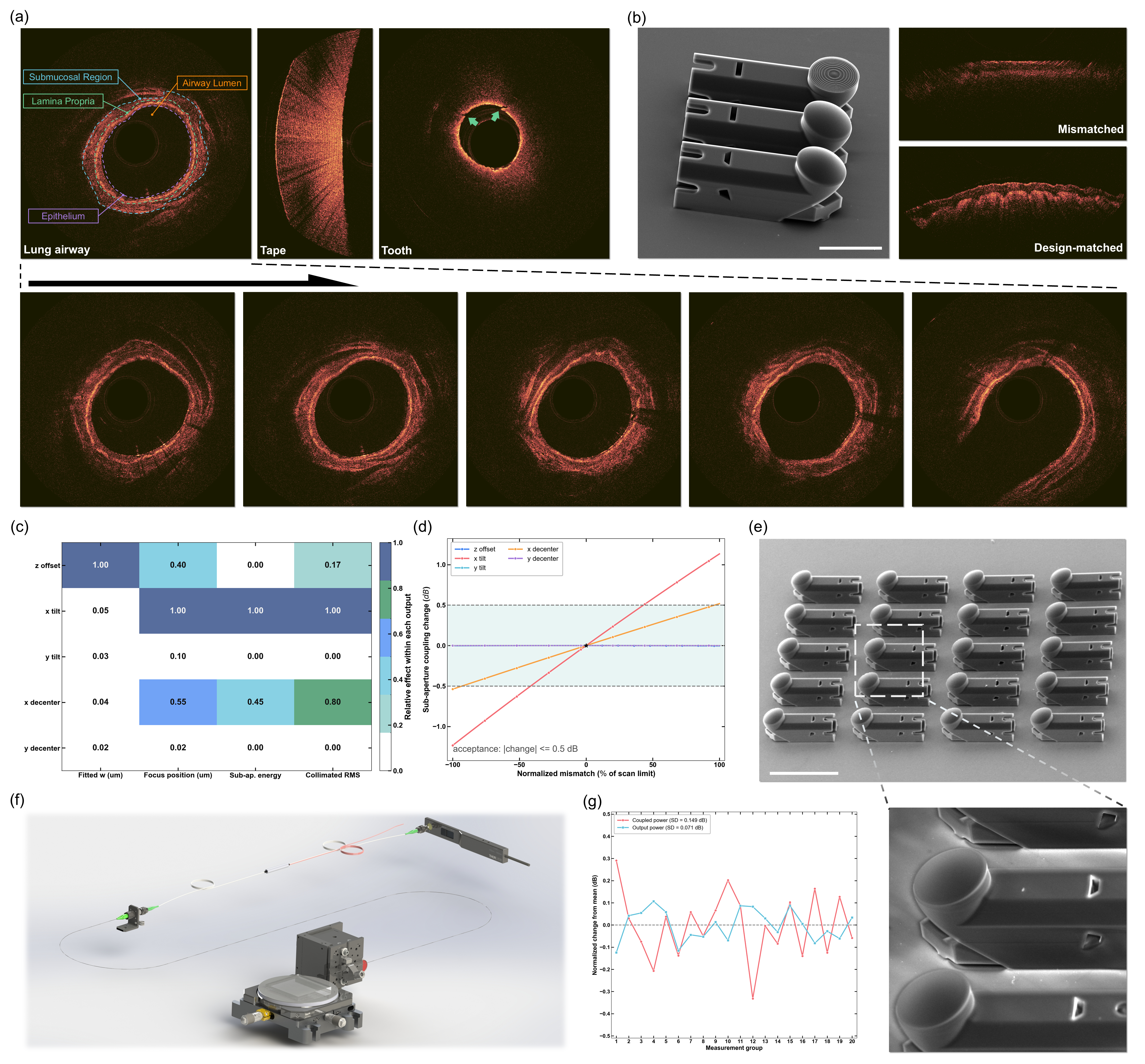}
\captionsetup{font=footnotesize}
\caption{\textbf{Application-adaptable imaging, assembly tolerance, and reproducibility of F2I-OCT probes.}
(a) Representative OCT imaging with F2I-OCT probes. Top, cross-sectional images of an ex vivo ovine airway, a wound adhesive tape roll, and an ex vivo human dental pulp cavity. Airway annotations indicate wall regions consistent with the epithelium, lamina propria, and submucosal region. Green arrows in the tooth image indicate fissure-like features along the cavity wall. Bottom, five representative airway B-scans selected from successive positions during rotational pullback; the arrow indicates the pullback direction. The complete pullback sequence and volumetric visualization are provided in Supplementary Videos~1 and~2.
(b) Representative interchangeable terminal-optic designs. Left, scanning electron micrograph of three printed probes, from bottom to top: a refractive objective designed for a nominal $1~\mathrm{mm}$ working distance beyond the outer tubing wall in air, a refractive objective designed for the corresponding working condition in water, and a diffractive-optical-element (DOE) terminal design. Scale bar, $200~\mu\mathrm{m}$. Right, OCT images of a fingertip acquired with the same water-optimized probe under design-matched (water) and medium-mismatched (air) conditions. Operation in air changes the refractive power of the curved tubing relative to the design condition, producing strong residual astigmatism, whereas immersion in water restores the designed refractive environment and substantially reduces the aberration.
(c) Numerical tolerance analysis of fiber--element misregistration. The heat map summarizes the relative influence of axial offset, tilt about the two transverse axes, and lateral decentre along the two transverse axes on the expanded-beam radius, focal position, power intercepted by the reference sub-aperture, and collimated-reference wavefront error. Each output is normalized to its largest response across the five degrees of freedom.
(d) Predicted change in coupled reference power as a function of normalized fiber--element misregistration. The shaded region marks the $\pm0.5~\mathrm{dB}$ acceptance window used for reference-guided assembly.
(e) Twenty optical elements fabricated in a single unattended array print before fiber assembly. The enlarged region marked by the dashed box shows representative terminal surfaces and their as-fabricated surface morphology. The array was imaged without conductive coating. Scale bar, $500~\mu\mathrm{m}$.
(f) Reference-guided bonding setup. A 1310~nm laser is launched through a fiber circulator into the delivery fiber, and the returned reference power is monitored in real time while the printed element is aligned. The adhesive is UV-cured once the return lies within the prescribed acceptance window.
(g) Returned reference power and side-viewing output power for twenty consecutively assembled probes, normalized to their respective set means. The standard deviations are $0.149~\mathrm{dB}$ and $0.071~\mathrm{dB}$, respectively.}
\label{fig:results}
\end{figure}

System-level imaging performance was first evaluated using assembled F2I-OCT probes. The conservatively estimated sensitivity was $93.1~\mathrm{dB}$, obtained from a measured signal-to-noise ratio of $62.5~\mathrm{dB}$ at a buried NOA~63--fused-silica interface with an estimated reflectance of $-30.6~\mathrm{dB}$, as described in Section~\ref{sec:experiments}. Representative depth-resolved images are shown in Fig.~\ref{fig:results}a. In a wound adhesive tape roll with a total thickness of approximately $2~\mathrm{mm}$, individual layers remained resolved across the full thickness. In an ex vivo ovine airway, the probe delineated the lumen and layered wall structures consistent with the epithelium, lamina propria, and submucosal region observed in endobronchial OCT\cite{berigei2024microscopic,yuan2022direct}. Imaging of an ex vivo dental pulp cavity similarly resolved the surrounding dental structure and revealed localized fissure-like features along the cavity wall, indicated by the green arrows. Five B-scans selected from successive pullback positions illustrate the evolution of airway morphology along the probe axis, while the complete rotational pullback and corresponding volumetric visualization are provided in Supplementary Videos~1 and~2.

The functional separation of the terminal imaging optics from the upstream expansion and reference modules allows the same F2I architecture to be adapted to different geometric and optical environments. Figure~\ref{fig:results}b shows three representative implementations. Two refractive objectives were designed for a nominal working distance of $1~\mathrm{mm}$ beyond the outer tubing wall but for different external media, air and water, respectively. A third probe incorporates a DOE as the terminal wavefront-forming element, demonstrating that the architecture is not restricted to conventional refractive or freeform surfaces. These designs share the same upstream beam-expansion and common-path reference architecture, while the terminal optics are independently adapted to the required working distance, surrounding medium, and wavefront transformation.

The influence of the surrounding medium is particularly important for a side-viewing probe because the curved tubing introduces different optical powers in the two principal planes, producing astigmatism that depends strongly on the refractive-index contrast at the tubing boundary\cite{li2018twophoton,li2020ultrathin}. This effect is illustrated in Fig.~\ref{fig:results}b using a probe whose terminal objective was optimized for operation in water. When the same probe is operated in air, the increased refractive-index contrast at the outer tubing surface changes its astigmatic contribution from the design condition, producing a pronounced separation of the orthogonal focal components and substantial degradation of the OCT image. Immersion in water restores the refractive environment for which the objective was optimized, substantially reducing the residual astigmatism and recovering the intended image quality. This comparison illustrates why working distance and surrounding medium must be incorporated into the terminal optical design rather than treated as external operating parameters. Together with the DOE implementation, it demonstrates that the terminal function can be adapted to different geometric, environmental, and wavefront-shaping requirements while the upstream F2I architecture remains unchanged. Probes incorporating these different terminal designs were operated on the same OCT platform without probe-specific adjustment of the reference path, dispersion, or polarization, demonstrating that application-specific terminal optics can be exchanged without reconfiguring the upstream interferometric system.

To determine whether this functional modularity could be retained after fiber assembly, the sensitivity of the printed architecture to fiber--element misregistration was evaluated numerically. Five degrees of freedom were varied independently over representative assembly tolerances: axial displacement from $0$ to $1~\mu\mathrm{m}$, tilt about each transverse axis from $-0.3^{\circ}$ to $+0.3^{\circ}$, and lateral decentre along each transverse axis from $-0.5$ to $+0.5~\mu\mathrm{m}$. Figure~\ref{fig:results}c summarizes their relative effects on four downstream quantities: expanded-beam radius, focal position, power intercepted by the reference sub-aperture, and wavefront error of the collimated reference field. The expanded beam and focal position are comparatively insensitive to the investigated perturbations, whereas displacement of the field across the wavefront-dividing TIR surface produces a stronger change in the intercepted reference power. The dominant sensitivity occurs for misalignment along the direction in which the sub-aperture is laterally displaced.

This behaviour is quantified in Fig.~\ref{fig:results}d. Over the investigated tolerance range, tilt in the sensitive transverse direction produces the largest change in coupled reference power, followed by lateral decentre along the same direction, whereas the orthogonal misalignments and axial displacement have substantially weaker effects. The $\pm0.5~\mathrm{dB}$ region therefore provides a convenient optical acceptance criterion for assembly. This sensitivity, rather than being solely a fabrication constraint, makes the embedded reference an effective indicator of the fiber--element registration that most strongly affects the interferometric state of the completed probe.

The fabrication route separates parallel production of the printed elements from final fiber assembly. Twenty elements were fabricated in a single unattended array print of approximately $12.5~\mathrm{h}$ (Fig.~\ref{fig:results}e). The enlarged scanning electron micrograph shows representative terminal surfaces within the array and confirms formation of the designed surface structures without gross defects. Because the beam expansion, reference path, and terminal optics are contained within each printed element, the remaining probe-specific fabrication step is the alignment and bonding of the element to a cleaved delivery fiber.

The reference sensitivity identified by the tolerance analysis is used directly to guide this bonding step. During assembly, 1310~nm light is launched through a circulator into the delivery fiber and the returned reference power is monitored continuously on a power meter (Fig.~\ref{fig:results}f). The printed element is first positioned on the fiber facet and then adjusted until the return lies within the prescribed acceptance range, after which the adhesive is UV-cured while the alignment is maintained. Because the feedback signal is generated by the embedded reference itself, the procedure does not require the swept-source OCT engine, interferometer, detector, or acquisition electronics. The same optical quantity that later provides the heterodyne reference therefore also closes the alignment loop during probe fabrication.

The twenty elements were subsequently assembled using this procedure and characterized consecutively. The returned reference power exhibited a standard deviation of $0.149~\mathrm{dB}$, while the side-viewing output power exhibited a standard deviation of $0.071~\mathrm{dB}$ across the twenty probes (Fig.~\ref{fig:results}g). The corresponding mean powers were $24~\mu\mathrm{W}$ for the returned reference and $10.5~\mathrm{mW}$ for the side-viewing output, with an average input power of $14~\mathrm{mW}$ measured at the output of the rotary joint. All reported powers are time-averaged values and therefore include the duty cycle of the swept source. The larger variation of the reference return is consistent with the tolerance analysis, which identifies reference coupling as the quantity most sensitive to residual fiber--element misregistration. Nevertheless, all twenty probes operated within the same acquisition window and required no probe-specific adjustment of the OCT interferometer.

Taken together, Fig.~\ref{fig:results} demonstrates two complementary consequences of the F2I architecture. The terminal imaging function can be redesigned for different working distances, surrounding media, and wavefront implementations without changing the upstream expansion or reference modules, while repeated fabrication and assembly preserve the optical state required for operation on the same OCT system. The architecture therefore supports interchangeability both across functionally different probe designs and across nominally identical probes produced in batch.

\section{Experiments}\label{sec:experiments} 

\subsection*{OCT System and Image Acquisition}

Imaging was performed with a swept-source OCT system based on a 1300~nm MEMS-VCSEL engine (SVM13X, Thorlabs) operated at 200~kHz, with a $\sim100$~nm $-15~\mathrm{dB}$ tuning range and an axial resolution of approximately $6~\mu\mathrm{m}$ in tissue. The photodetector signal was high-pass filtered (SHP-20+, Mini-Circuits) and digitized at 1.0~GS/s with 12-bit resolution (ATS9371, Alazar Technologies), together with the rotary-encoder signal.

Side-view scanning was performed using a fiber-optic rotary joint (MJP-FAPB-131-28-FA, Princetel) driven by a brushless motor (EC-max 30 with ESCON 50/5, maxon) and combined with a motorized pullback stage (Z812B/TDC001, Thorlabs). The rotary joint operated at 3000~rpm. A 512-pulse-per-revolution motor encoder and a 2:1 transmission provided 1024 angular references per probe revolution, from which 1024-A-line B-scans were reconstructed by angular resampling. Data acquisition, GPU processing, and real-time visualization were implemented in C++/CUDA/Qt.

The same OCT system was used for all probe designs without probe-specific adjustment of the reference path, dispersion, or polarization. For the dual-arm comparison in Fig.~\ref{fig:commonPath}f, an otherwise identical distal imaging design without the printed reference module was operated in a Michelson configuration. The reference arm was independently path matched, and neither configuration used dispersion compensation.

\subsection*{Optical Design and Numerical Modeling}

Optical calculations used a polymer refractive index of 1.498 at 1310~nm. The in-element beam-expansion and reference modules were modeled by angular-spectrum propagation on a $1024\times1024$ computational grid. The expanded field was independently evaluated using CODE~V beam synthesis. Fiber--element misregistration was analyzed using the same angular-spectrum framework. Axial displacement was varied from $0$ to $1~\mu\mathrm{m}$, tilt about each transverse axis from $-0.3^{\circ}$ to $+0.3^{\circ}$, and lateral decentre along each transverse axis from $-0.5$ to $+0.5~\mu\mathrm{m}$. The resulting changes in expanded-beam radius, focal position, reference-sub-aperture power, reference wavefront error, and coupled reference power were evaluated. Complete tolerance curves and numerical details are provided in the Supporting Information.

Terminal imaging optics were designed in OpticStudio (Ansys Zemax) using physical-optics propagation through the catheter sheath. Refractive terminal designs were optimized according to the required working distance and surrounding medium, with astigmatism correction implemented using a custom user-defined surface, while the upstream beam-expansion and reference modules were kept unchanged. Diffractive terminal optics were designed separately and implemented as surface-relief profiles for two-photon grayscale lithography. Detailed surface prescriptions and additional simulation parameters are provided in the Supporting Information.

\subsection*{Two-Photon Fabrication and Probe Assembly}

Optical elements were fabricated using a Quantum X shape system (Nanoscribe) with IP-S photoresist on fused-silica substrates. The main probe structures and refractive terminal surfaces were fabricated by conventional two-photon polymerization (2PP) with $0.2~\mu\mathrm{m}$ slicing and hatching distances. Diffractive optical element (DOE) terminal surfaces were fabricated on the same platform using two-photon grayscale lithography (2GL), in which the local exposure dose was modulated to generate the designed surface-relief profile. A complete F2I-OCT element required approximately 35~min to print, and 20 elements were fabricated in a single unattended array print of approximately $12.5~\mathrm{h}$. After printing, the structures were developed in PGMEA for 45~min and rinsed in Novec~7100 for 2~min.

Each printed element was bonded to a cleaved single-mode fiber using a custom four-axis alignment stage. During reference-guided bonding, 1310~nm light was launched through a fiber circulator and the returned reference power was continuously monitored with an optical power meter. The element was first positioned on the fiber facet and subsequently adjusted until the reference return fell within the prescribed acceptance range. The adhesive was then UV-cured while this condition was maintained.

\subsection*{Optical Characterization and Reproducibility}

Expanded beam profiles were measured using an InGaAs camera (Ninox 640 II, Raptor Photonics) through a 20$\times$ NIR objective (M Plan Apo NIR, Mitutoyo) and a 150~mm tube lens, corresponding to $1~\mu\mathrm{m}$ per pixel in object space. Beam profiles were measured over a $0.7~\mathrm{mm}$ axial range and fitted with rotated two-dimensional Gaussian functions to determine the beam radius and propagation characteristics.

Optical powers were measured using a PM100D meter with an S132C sensor (Thorlabs). For the probe-reproducibility measurements, returned reference power and side-viewing output power were recorded consecutively for the twenty assembled probes under the same source conditions. Reported optical powers are time-averaged values and therefore include the duty cycle of the swept source. Probe-to-probe variations were quantified relative to the corresponding set means.

Sensitivity was measured using the buried interface between cured NOA~63 optical adhesive and fused silica. Because a reliable refractive index for cured NOA~63 at 1310~nm was unavailable, a conservative working estimate of $n=1.535$ was used to calculate the Fresnel reflectance of the interface. The signal-to-noise ratio was calculated from 2048 A-lines as
\begin{equation}
    \mathrm{SNR}
    =20\log_{10}
    \left(
    \frac{A_{\mathrm{peak}}}{\sigma_{\mathrm{noise}}}
    \right),
\end{equation}
where $\sigma_{\mathrm{noise}}$ is the standard deviation over a signal-free depth range. The sensitivity was calculated as
\begin{equation}
    S=\mathrm{SNR}-R_{\mathrm{dB}},
\end{equation}
where $R_{\mathrm{dB}}$ is the calculated Fresnel reflectance of the test interface in decibels.

Printed structures were examined by scanning electron microscopy. Unless otherwise noted, samples were sputter-coated with gold before imaging. The batch-fabricated array shown in Fig.~\ref{fig:results}e was imaged without conductive coating to preserve the as-fabricated structures. Surface figure of the refractive terminal objective designed for operation in air was characterized by coherence-scanning interferometry (NewView 8300, Zygo) using a 20$\times$ Mirau objective. Measurements were performed over representative sub-apertures sampling the optical surface, and the corresponding measurement geometry and surface-error maps are provided in the Supporting Information.

\subsection*{Sample Preparation}

Extracted human teeth were used for dental pulp cavity imaging under Institutional Review Board protocol STUDY00006543. The crowns were mechanically prepared to expose the internal pulp cavities while preserving the surrounding dentin. During imaging, the opened cavity was positioned to provide access for the side-viewing probe.

Fresh ovine lung tissue was used for airway imaging. Immediately after collection, the tissue was maintained in phenol red-free Dulbecco's Modified Eagle Medium (DMEM) to preserve hydration and tissue integrity. The samples were transported on dry ice by overnight shipment and prepared for OCT imaging after receipt.

\section{Discussion and conclusion}\label{sec:conclusion}

The central distinction of F2I-OCT is the use of two-photon microfabrication to define the complete distal interferometric architecture rather than only the terminal imaging optic. Conventional dual-arm probes retain direct control of the reference field but require a separate interferometer arm and probe-specific matching, whereas interface-based common-path probes simplify the system while coupling the reference level and delay to the optical interface from which the reference is generated\cite{gora2017endoscopic,sharma2007common,park2012double}. Previous 2PP endoscopic probes have greatly expanded the design freedom of distal imaging optics, including beam redirection, aberration correction, and wavefront shaping, but beam expansion and interferometric reference generation have generally remained distributed among separately processed fiber sections or external system components\cite{li2018twophoton,li2020ultrathin,ruchka2025printed}. F2I-OCT brings these functions into a single fiber-mounted printed body. Beam expansion, side-view redirection, common-path reference generation, and terminal imaging are physically integrated while their principal optical parameters remain independently designable. The architecture is therefore monolithic in implementation without becoming monolithic in function.

Encoding beam expansion and reference generation directly in the printed geometry changes both how the probe is fabricated and how it interfaces with the OCT system. The in-element expander replaces fusion-spliced and length-controlled fiber sections used to establish the distal beam, while the wavefront-divided reference makes the reference strength and optical delay intrinsic properties of the probe rather than quantities that must be re-established after probe exchange. Printing the optical elements independently of the delivery fibers enables parallel fabrication, and the embedded reference itself provides a sensitive optical signal for guiding the final fiber--element bond. Across twenty consecutively assembled probes, the returned-reference and side-viewing-output powers showed standard deviations of $0.149~\mathrm{dB}$ and $0.071~\mathrm{dB}$, respectively, while all probes operated within the same OCT acquisition configuration without probe-specific adjustment of the reference path, dispersion, or polarization. Probe replacement therefore becomes an exchange of a self-contained interferometric front end rather than a new interferometer-matching procedure.

Because the expansion and reference modules are established upstream of the terminal optics, the imaging function can be redesigned for different working distances, surrounding media, sheath conditions, and wavefront requirements without re-optimizing the rest of the architecture. The air- and water-optimized refractive probes demonstrated here illustrate this flexibility in two qualitatively different optical environments. Airway imaging operates in an air-filled lumen, whereas intravascular OCT is typically performed in a liquid-filled environment during saline or contrast flushing; for a curved catheter sheath, these environments impose different refractive powers and therefore different astigmatic corrections on the distal optics\cite{gora2017endoscopic,yuan2022direct,zhang2024pneumaoct}. The DOE implementation further shows that the terminal function is not restricted to conventional refractive or freeform surfaces. These design freedoms are relevant beyond airway and intravascular imaging: gastrointestinal and fallopian-tube probes must operate within different lumen dimensions and access geometries\cite{li2019capsule,keenan2017falloposcope}, image-guided biopsy and intervention benefit from probes that can be exchanged or customized without reconfiguring the imaging system\cite{wang2024automatic}, dental and root-canal imaging require access to narrow cavities and sensitivity to internal interfaces and structural defects\cite{shemesh2007ability,alkhani2026optical}, and miniature neuroimaging probes impose particularly stringent constraints on probe dimensions and working distance\cite{xu2024ultrathin}. In this framework, application specificity can be introduced primarily through the terminal optical design rather than by rebuilding the upstream interferometric architecture.

Reference strength and optical delay provide an additional set of design variables that can be adapted independently of the terminal imaging optics. The wavefront-dividing sub-aperture allows the reference level to be selected according to the expected sample return, while the folded internal path determines the zero-delay position and can be modified according to the required imaging range. Because this optical path difference is encoded directly in the printed geometry rather than established through probe-specific mechanical adjustment, the same concept may also be useful outside biological endoscopy. OCT has already been explored for robotic inspection of inaccessible engineered structures and for industrial dimensional and defect inspection\cite{he2023robotic,fu2024progress,kahatapitiya2026optical}. In such applications, interchangeable probes with geometry-defined axial coordinates could be combined with terminal optics adapted to different bore dimensions, stand-off distances, refractive environments, or target reflectivities. The metrological advantage of such a geometry-defined reference has not been quantitatively established here, but it provides a concrete direction in which the architecture could extend from imaging toward reproducible confined-space measurement.

The present study establishes the optical architecture, fabrication strategy, probe reproducibility, interchangeability, and structural-imaging capability of F2I-OCT across representative ex vivo and benchtop demonstrations. This architecture provides a foundation for further translation toward diagnostic, interventional, and industrial applications, where future studies can evaluate application-specific performance through larger biological sample sets, histological or procedural validation, in vivo imaging, or quantitative confined-space measurements. The same platform can be specialized at the device level by redesigning working distance, numerical aperture, immersion correction, refractive or diffractive wavefront shaping, reference strength, and reference delay without changing the underlying system concept. Advanced terminal optics can further tailor focal structure and depth response within fiber-scale OCT probes\cite{qiu2020uniform,xi2014diffractive,gu2024computer,he2025planar}, while alternative printable materials could broaden the accessible refractive-index range, thermal stability, and environmental compatibility\cite{hong2021three,ye2024solventfree,you2025extremely}. More broadly, F2I-OCT extends the role of additive microfabrication from producing increasingly sophisticated individual micro-optics to defining complete fiber-scale optical architectures. Refractive, reflective, diffractive, beam-dividing, interferometric, and focusing functions can be encoded within a common three-dimensional structure so that additional optical functionality is introduced through design rather than through a proportional increase in discrete components, fiber-processing steps, and alignment operations. Two-photon printing can therefore serve not only to miniaturize optical components, but also to integrate and simplify complete fiber-scale optical systems.

\section*{Acknowledgments}

The authors thank Dr. Kathleen Vincent for providing the ovine tissue used in this study. This work was supported by the National Institutes of Health through the Office of the Director (S10OD036299), the National Cancer Institute (R21CA277667 and R21CA268190), and the National Institute of Biomedical Imaging and Bioengineering (R01EB037023).

\section*{Ethics Statement}

Extracted human teeth were used under Institutional Review Board protocol STUDY00006543.

\section*{Conflicts of Interest}

R. Liang is the founder of Light Research Inc. and LR InnovOptics Inc. These relationships have been disclosed to and are managed by the University of Arizona in accordance with its conflict-of-interest policies. The remaining authors declare no conflicts of interest.

\section*{Data Availability Statement}

The data supporting the findings of this study are available within the article and its Supporting Information, including the user-defined-surface source code. Additional data are available from the corresponding author upon reasonable request.

\bibliographystyle{unsrt}
\bibliography{wileyNJD-Chicago}

\section*{Supporting Information}

Additional supporting information can be found online in the Supporting Information
section.


\end{document}